\documentclass[12pt,letterpaper]{article}
\usepackage{t1enc}
\usepackage[latin1]{inputenc}
\usepackage[english]{babel}
\begin{document}

\title{An Approach to the Statistics of Turbulence}
\author{Edsel A. Ammons \\ Department of Chemistry and Physics \\ Arkansas State University \\ P. O. Box 419 \\ State University, AR 72467}
\maketitle
\abstract{\vskip .250in}
A calculational approach in fluid turbulence is presented.  A turbulent environment is thought to be a condition of the system for which the time and space evolution of the dynamical variables are chaotic, but for which a statistical distribution function is applicable.  The time-evolution of an initially given probability distribution function for the dynamical variables is given by the deterministic evolution equations that govern the system, called deterministic chaos.  A fluid environment is a dynamical system, having energy input mechanisms at long distance scales occurring on the spatial boundaries and energy dissipation mechanisms due to viscosity occurring on small distance scales.  Hence, the fluid has an attractor, called the Strange Attractor in the turbulent regime, for its dynamical variables.  It is proposed here that the fluid probability density functional also has an attractor for its time-evolution, and an approach to generating this time-evolution of the statistics is presented.  The same mechanism that causes the dynamical variables to have an attractor in phase space, that is the tendency for the equilibration of energy input rates and energy output rates to set in, also causes an arbitrary initial statistics to evolve toward an attracting statistics, which is stationary in time.  It is this stationary statistics that allow the Kolmogorov scaling ideas to have applicability.  The evolution of the fluid's statistics can be set up as part of an approximate space-time path integral.  Ensemble averages of any dynamical variable can be formulated in terms of this path integral.  Fluid space-time configuration sampling techniques naturally suggest a useful way, using a relatively arbitrary initial statistics functional to calculate averages.

\section{Description of the Mathematical Approach to Fluid Turbulence}
Let us set up the evolution equations for incompressible fluid dynamics.  The extension of the proposal to be described to compressible fluid dynamics will pose no unusual difficulty.

We have, $$\rho \frac{d \vec{v}}{dt}=\vec{f}+\eta \nabla ^{2} \vec{v},$$ where $\vec{f}$ is an external force density, such as due to a scalar pressure field, and $\eta $ is the coefficient of viscosity.  $\vec{v}$ is the fluid velocity field.  We also have, $$\nabla \cdot (\rho \vec{v}) + \frac{\partial{\rho}}{\partial{t}}=0.$$  Here, $\rho $ is the mass density of the fluid.  If this mass density is a constant, then the velocity field is divergenceless.  Also, $$ \frac{d \vec{v}}{dt} =\frac{\partial \vec{v}}{\partial{t}}+\vec{v} \cdot \nabla \vec{v}.$$  So we have the fluid dynamic system, \begin{eqnarray} \frac{\partial \vec{v}}{\partial t} & = & -\frac{\nabla P}{\rho }-\vec{v} \cdot \nabla \vec{v} + \nu \nabla^{2} \vec{v} \label{eq:Dynamics} \\ \nabla \cdot \vec{v} & = & 0, \label{eq:Navier-Stokes}
\end{eqnarray} where $P$ is the pressure, and $\nu \equiv \frac{\eta }{\rho }$.

We drop the external force density in what follows.  This is not an essential step.  What are needed are a set of interacting fields, not restricted to being velocity fields, together with stationary boundary conditions to allow the deterministic time-evolution of the set.  We also associate with the velocity field a probability density functional, $\rho [v,t].$  The fluid statistics time-evolves according to deterministic chaos \cite{bib:Rosen} \cite{bib:Thacker},  $$\rho [v_{f},t_{f}] = \int d[v_{0}]K[v_{f},t_{f};v_{0},t_{0}] \rho [v_{0},t_{0}],$$ where the kernel is, $$K[v,t;v_{0},t_{0}]=\delta [v_{f} - v_{classical}[t_{f};v_{0},t_{0}]].$$  That is, the number, $\rho $, associated with the velocity field $v_{0}$ at time $t_{0}$ will be associated with $v_{f}$ at time $t_{f}$, where $v_{f}$ is the velocity field $v_{0}$ deterministically evolves into from time $t_{0}$.  Given a functional of the spatial velocity field, $A[v]$, its ensemble average, at time $t_{f}$ is, $$<A[v]>=\int d[v_{f}]A[v_{f}]K[v_{f},t_{f}; v_{0},t_{0}] \rho[v_{0},t_{0}]d[v_{0}].$$

We want to propagate the fluid's statistics according to deterministic chaos, even though the detailed fluid orbits are chaotic.  Let, \begin{eqnarray*} <A[v]> & = & \int A[v_{f}] \rho [v_{f},t_{f}]d[v_{f}] \\ &= & \int A[v_{f}]K[v_{f},v_{0}]\rho[v_{0},t_{0}]d[v_{f}]d[v_{0}] \\ & = & \int A[f[v_{0}]] \rho [v_{0},t_{0}]d[v_{0}], \end{eqnarray*} where, \begin{eqnarray*} K[v_{f},v_{0}] &=& \delta [v_{f}-f[v_{0}]]\\ & =& \int \delta [v_{f} - f_{1}[v_{1}]]\delta [v_{1}-f_{1}[v_{0}]]d[v_{1}] \\ & =& \int \delta [v_{f}-f_{2}[v_{1}]] \delta [v_{1}-f_{2}[v_{2}]] \delta [v_{2}-f_{2}[v_{0}]]d[v_{1}]d[v_{2}] \\ & = &\int \delta [v_{f}-f_{3}[v_{1}]] \delta [v_{1}-f_{3}[v_{2}]] \delta [v_{2}-f_{3}[v_{3}]] \delta [v_{3}-f_{3}[v_{0}]]d[v_{1}]d[v_{2}]d[v_{3}] \\ & = &\ldots, \end{eqnarray*} where the velocity fields, $v_{1}$, $v_{2}$, $v_{3}$, etc. occur in chronological order, $v_{1}$ being closest to time $t_{f}$.  Eventually, we have an $f_{M}$, where $M$ is large, such that $v_{M}=f_{M}[v_{0}]$ is infinitesimally different from $v_{0}$.

So, $$<A[v]>=\int d[v] A[v_{f}] \delta [v-f_{M}[v]] \rho [v_{0},t_{0}],$$ where, $$d[v]=d[v_{f}]d[v_{1}] \cdots d[v_{M}]d[v_{0}],$$ and, $$<A[v]>=\int A[f_{M}[ \cdots [f_{M}[v_{0}]] \cdots ] \rho [v_{0},t_{0}]d[v_{0}].$$

We must have, $$K[v_{f},v_{0}]=\delta [v_{f}-f[v_{0}]]=\delta [f^{-1}[v_{f}]-v_{0}].$$  We have, \begin{eqnarray*} \rho [v_{f}, t_{f}] & = & \int \delta [f^{-1}[v_{f}]-v_{0}]\rho [v_{0}, t_{0}]d[v_{0}] \\ & = & \rho [f^{-1}[v_{f}], t_{0}] \\ & = & \rho [v_{0},t_{0}]. \end{eqnarray*}  The exact rule, $f_{M}[v]$, requires a solution of the fluid dynamic equations, incorporating boundary conditions.  The exact rule, $f_{M}[v]$, is difficult to find.  Let us use an approximate rule,  $F_{M}[v].$  

Then we may say, with motivation to follow, \begin{eqnarray} <A[v]> & = \int d[v]A[v_{f}] \delta[v-F_{M}[v]]\lambda [\nabla \cdot v] \lambda [v - v_{B}] \rho [v_{0}, t_{0}]. \end{eqnarray}  The functional integration is over all space-time velocity fields within the spatial system, between times $t_{0}$ and $t_{f}.$  $\delta [v-F_{M}[v]]$ is a space-time delta functional.  $F_{M}[v]$ generates $v$ from a $v$ an infinitesimal instant earlier.  The $\lambda $ functionals are evaluated at a particular instant.  The needed properties of the $\lambda $ functionals are $\lambda [0]=1$, and $\lambda [g \neq 0]=0.$  $\lambda [g]$ could be, $$\lim_{\epsilon_{1}\rightarrow 0^{+}} e^{-g^{2}/\epsilon_{1}}.$$  We have, then, \begin{eqnarray} <A[v]> & = & \int d[v_{0}]A[F_{M}[F_{M}[ \cdots [F_{M}[v_{0}]] \cdots ] \rho [v_{0},t_{0}] \nonumber \\ & & \lambda [F_{M} [ \cdots [F_{M}[v_{0}]] \cdots ] - v_{B}] \cdots \lambda [F_{M}[v_{0}] - v_{B}] \lambda [v_{0} - v_{B}] \nonumber \\ & & \lambda [\nabla \cdot F_{M} [ \cdots [F_{M}[v_{0}]] \cdots ]] \cdots \lambda [\nabla \cdot F_{M} [ v_{0}]] \lambda [ \nabla \cdot v_{0}]. \label{eq:AAverage} \end{eqnarray}  The right-hand side of (\ref{eq:AAverage}) equals $<A[v]>,$ because $\rho [v_{0},t_{0}]$ will be non-zero only for $ v_{0}$'s that make all the $\lambda $'s equal one.  This means that, ideally, we utilize only spatial velocity fields that are divergenceless, and that satisfy the stationary boundary conditions.

The velocity field has an attractor, determined by the stationary boundary conditions on the fluid.  When the boundary conditions allow steady laminar flow to become established, the attractor consists of a single velocity field.  When the Reynolds number becomes large enough, bifurcations set in, or the onset of instability occurs, and the attractor begins to consist of more than one spatial velocity field.  In the turbulent regime, the attractor consists of many velocity fields, and the fluid accesses these velocity fields according to a probabilty distribution.

Given a functional of the spatial velocity field, $A[v]$, and the fluid dynamic system of equations (\ref{eq:Dynamics}), (\ref{eq:Navier-Stokes}), we will say that its ensemble average when the system has reached its attractor is, \begin{eqnarray} <A[v]> & = & \lim_{t_{f} - t_{0} \rightarrow \infty } \int d[v]A[v_{f}] \delta [v-F_{M}[v]] \lambda [\nabla \cdot v] \lambda [v-v_{B}] \nonumber \\ & & \cdot  \rho [v_{0},t_{0}].
 \label{eq:PathIntegral}  \end{eqnarray}
The delta functional condition, $\delta [v-F_{M}[v]]$ implements equation (\ref{eq:Dynamics}).  $F_{M}[v]$ is an approximate rule for carrying $v$ from an earlier instant to a later instant.  In the first approximation, $$F_{M}[v] = v - \Delta t ( \vec{v} \cdot \nabla v -  \nu \nabla^{2}v),$$ where $v$ is evaluated at an instant $\Delta t$ earlier.  $\lambda [\nabla \cdot v]$ implements a zero divergence condition on the spatial velocity fields, and $\lambda [v-v_{B}]$ requires the spatial velocity fields to have values $v_{B}$ on the spatial boundaries.

Let us consider the path integral (\ref{eq:PathIntegral}) to be on a space-time lattice.  We could use $$\delta (x) = \lim_{\epsilon \rightarrow 0^{+}} \frac{e^{-\frac{x^{2}}{\epsilon }}}{ \sqrt{ \pi \epsilon}}.$$  We have for the average of $A[v]$ in the steady-state (attractor), letting $\epsilon_{1}(\Delta x )^{2}=\epsilon$, where $\epsilon_{1}$ is in the $\lambda$ functional for the zero divergence condition, \begin{eqnarray} <A[v]> & = & \lim_{ \epsilon \rightarrow 0^{+}} \lim_{t_{f} - t_{0} \rightarrow \infty } \int d[v]e^{-H[v]/ \epsilon } A[v_{f}] \rho [v_{0}]. \label{eq:Formula} \end{eqnarray}

Also, $$\rho [v_{0}] \equiv \rho [v_{0}, t_{0}],$$ and the space-time integration measure, $d[v]$, is with boundary effects neglected, $$d[v]=(\frac{1}{\sqrt{\pi \epsilon }})^{3N} \prod_{ijkl}dv_{ijkl}.$$  $H[v]$ is a functional of the lattice space-time velocity field.  Also, neglecting boundary effects, $$H[v] = \sum ((v_{l} - v_{l-1} - g[v_{l-1}]\Delta t)^{2} + (v_{x,ijkl} - v_{x, i-1,jkl} + \cdots )^{2}) + \sum'(v_{ijkl} - v_{B,ijkl})^{2},$$ $$g[v_{l}]= - \vec{v} \cdot \nabla v_{l} + \nu \nabla^{2}v_{l}.$$  Or, $$g[v_{l}] = -v_{x,ijkl} \frac{(v_{ijkl} - v_{i-1,jkl})}{\Delta x} + \cdots + \nu \frac{(v_{ijkl} - v_{i-1,jkl} - v_{i-1,jkl} +v_{i-2,jkl})}{(\Delta x)^{2}} + \cdots .$$  $N$ is the number of space-time lattice points in the system.  We have $\sum $  as a sum over all space-time lattice points in the system, neglecting boundary effects, and $\sum^{'}$ as a sum over all space-time lattice points on the spatial boundary.  Also, $ijk$ are spatial lattice indicies, and $l$ is the index in the time direction.

This discretization technique is expected to get better as one increases the lattice fineness and makes use of higher order finite difference approximations applied to partial differential equations \cite{bib:Warsi}.  A good approximation to the attracting statistics as a starting point will shorten the evolution time required for accurate results for averages occurring in the steady state.  The path integral (\ref{eq:Formula}) can be evaluated with Monte Carlo techniques utilizing importance sampling.  A calculation of the stationary velocity field that would exist, for the given boundary conditions, if that field were stable, should be a good starting point from which to begin a sampling of the velocity field space-time configuration space.

We have, $$<A[v]> = J\sum_{i=1}^{n}\frac{A[v_{f,i}] \rho [v_{0,i}]}{n},$$ where, $$ J = \int e^{-H[v]/\epsilon }d[v],$$ and the space-time configurations are distributed according to the weighting $e^{-H[v]/\epsilon }.$  $n$ is the number of space-time configurations in the ensemble.  $A[v_{f,i}]$ is the observable $A[v]$ evaluated at the final time slice of the $i^{th}$ space-time configuration.  The value $\rho [v_{0,i}]$ is attached to the initial time slice of the $i^{th}$ configuration.

We need, $$ 1 = J \sum_{i =1}^{n} \frac{\rho [v_{0,i}]}{n}.$$  So, we must do the integral, $J,$ and constrain $\sum \rho [v_{0,i}]$ to equal $n/J.$

Going to higher order finite difference approximations for the derivatives will eventually require consideration of space-time boundary effects for each space-time point in the system.  Even for the initial approximation, boundary considerations must be made for points near the boundaries.

\section{Summary}

We have said that the time evolution of the statistics also has an attractor for the following reasons; \begin{enumerate} \item It is a way to get a solution to the problem of arriving at the steady-state turbulence statistics. One knows that the steady state statistics is stationary with respect to its time-evolver.  Proposing that this statistics is the attractor of its time-evolver means one does not have to have the statistics to get the statistics, thereby offering an approach to the closure problem for the determination of correlations.  \item The statistical physics approach has been successful in equilibrium thermodynamics where the derivation of the microcanonical ensemble can be taken as the indication that equilibrium is the attractor for the dynamical system when the boundary conditions on the system input no energy.  In the attractor, the mean energy input is equilibrated with the mean energy output, because in the attractor dissipative losses have effectively shut off, and the system becomes effectively Hamiltonian.  The stationarity of the statistics requires the vanishing of the Poisson bracket of the statistics with this Hamiltonian resulting in the statistics of equal a priori probability. \item The dynamical system, of which a fluid is an example, has an attractor \cite{bib:Ruelle}.  The dynamics of the statistical approach should mirror the dynamics of the actual dynamical system.  \item The statistics of the dynamical system prior to reaching the attractor has no reason to be unique.  The statistics of the attractor is unique, in which the geometry of the system, the stationary boundary conditions, and the viscosities, all of which determine the Reynolds number, play a crucial role in determining the attractor.  \item The stationary statistics of the fluid occurs when the equilibration of energy input and energy output has set in \cite{bib:Frisch}.\end{enumerate}
 \section{Conclusions}

In the discretized version of the path integral that attempts to arrive at the stationary statistical effects in the generation of the ensemble average of a dynamical variable, using an approximate rule for the dynamics, one should arrive at, in the continuum limit, a greater insensitivity to the initial statistics and a generation of steady-state statistical effects.  One is trying to use the turbulent transience to get at steady-state turbulence effects.  These steady-state statistical effects become the backdrop for Kolmogorov's ideas of self-similarity and the resulting scaling exponents.

\section{Acknowledgments}

I wish to acknowledge the Department of Chemistry and Physics of the Arkansas State University, Jonesboro, for the environment necessary for this work.  I wish to thank Professor Leo P. Kadanoff and Professor Joseph A. Johnson for informative and useful discussions.

\end{document}